\documentclass[twoside,a4paper,11pt]{proceedings}
\usepackage{graphicx}
\usepackage{hyperref}
\usepackage{movie15}
\usepackage{natbib}
\usepackage{units}
\usepackage{textcomp}
\usepackage{multicol}

\usepackage{amsmath}	
\usepackage{amssymb}	
\usepackage{verbatim}
\usepackage[utf8]{inputenc}
\usepackage[dvipsnames]{xcolor}
\usepackage{colortbl}
\begin{document}
\pagenumbering{arabic}
\pagestyle{myheadings}
\thispagestyle{empty}
\vspace*{-1cm}
{\flushleft\includegraphics[width=3cm,viewport=0 -30 200 -20]{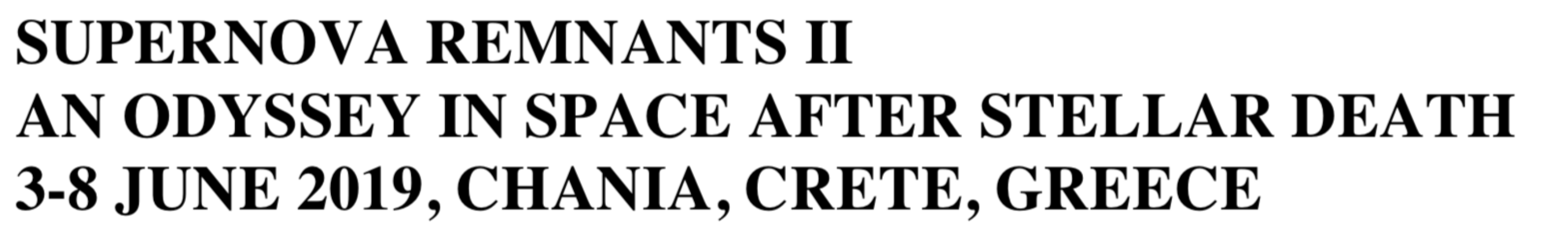}}
\vspace*{0.2cm}
\begin{flushleft}
{\bf {\LARGE
Dust destruction by the reverse shock in the clumpy supernova remnant Cassiopeia~A based on hydrodynamic simulations
}\\
\vspace*{1cm}
Florian Kirchschlager\footnote{E-mail: f.kirchschlager@ucl.ac.uk}, Franziska~D.~Schmidt$^{1}$, M.~J.~Barlow$^{1}$, Erica~L.~Fogerty$^{2}$, Antonia Bevan$^{1}$ and  Felix~D.~Priestley$^{1,3}$
%
}\\
\vspace*{0.5cm}
%
 $^{1}$Department of Physics and Astronomy, University College London, Gower Street, London WC1E 6BT, United Kingdom\\
 $^{2}$Center for Theoretical Astrophysics, Los Alamos National Lab, Los Alamos, NM 87545, United States\\
 $^{3}$School of Physics and Astronomy, Cardiff University, Queen's Buildings, The Parade, Cardiff, CF24 3AA, United Kingdom
%
\end{flushleft}
\markboth{
Dust destruction in Cas~A
}{
Kirchschlager et al.
}
\thispagestyle{empty}
\vspace*{0.4cm}
\begin{minipage}[l]{0.09\textwidth}
\ 
\end{minipage}
\begin{minipage}[r]{0.9\textwidth}
\vspace{1cm}
\section*{Abstract}{\small
Observations of the ejecta of core-collapse supernovae have shown that dust grains form in over-dense gas clumps in the expanding ejecta. The clumps are later subject to the passage of the reverse shock and a significant amount of the newly formed dust material can be destroyed due to the high temperatures and high velocities in the post-shock gas. To determine dust survival rates, we have performed a set of hydrodynamic simulations using the grid-based code  \textsc{\mbox{AstroBEAR}} in order to model a shock wave interacting with a clump of gas and dust. Afterwards, dust motions and dust destruction rates are computed using our newly developed external, post-processing code \textsc{\mbox{Paperboats}}, which includes gas and plasma drag, grain charging, kinematic and thermal sputtering as well as grain-grain collisions. We have determined  dust survival rates for the oxygen-rich supernova remnant Cassiopeia~A as a function of initial grain sizes, dust materials and clump gas densities.
\normalsize}
\end{minipage}

\section{Introduction}
 
It is well established that dust grains can form in over-dense gas clumps in the ejecta of supernova remnants (\mbox{SNR}; \citealt{Lucy1989}). On the other hand, a large fraction of the dust is potentially destroyed by the reverse shock which occurs when the supernova blast wave hits the circumstellar and interstellar material. Previous studies (\citealt{Nozawa2007, Bianchi2007,Nath2008,Silvia2010,Silvia2012,Bocchio2014, Biscaro2016,Micelotta2016}) determined the dust survival rate for a wide range of initial conditions and destruction processes. In order to investigate the motion and destruction of dust grains in a SNR on the basis of hydrodynamic simulations -- taking into account comprehensive dust physics -- we have developed the post-processing code \textsc{\mbox{Paperboats}} (\citealt{Kirchschlager2019b}). Unlike many other studies, both sputtering \textit{and} grain-grain collisions are considered as destruction processes, providing a more complete picture of the dust evolution in SNRs. 

The dusty remnant Cassiopeia~A (Cas~A) with an age of ${\sim}\,350$ years (\citealt{Kamper1976}) provides a unique laboratory to investigate the dust destruction by a reverse shock.  Cas~A is oxygen-rich (\citealt{Chevalier1979}) and has a highly clumped structure (\citealt{Milisavljevic2013}).  The dust survival rate of remnants like Cas~A is crucial for determining whether supernovae significantly contribute to the dust budget in the interstellar medium.

\section{Hydrodynamic simulations}

In order to simulate the dynamical evolution of a SNR reverse shock impacting the clumpy ejecta, we used the grid-based hydrodynamic code \textsc{\mbox{AstroBEAR}} (\citealt{Carroll-Nellenback2013}).  We pursue the cloud-crushing scenario (\citealt{Woodward1976,Silvia2010}) in which a planar shock is driven into a single clump of gas that is embedded in a low-density gaseous medium. The results of this representative clump can be projected then to the entire remnant. The clump destruction depends strongly on the ratio $\chi$ between the gas densities of the clump and the ambient medium. The simulation is executed for three cloud-crushing times (\citealt{Klein1994}) after the first contact of the shock with the clump which amounts to $\unit[{\sim}\,60]{yr}$ years for the density contrast $\chi=100$. The size of the domain is chosen to ensure that the dust does not flow out of the domain at the back end during the simulation time. We consider\break here only 2D simulations due to the large computational effort for highly resolved 3D post-processing simulations. Radiative cooling is taken into account for a gas of pure oxygen.

\section{Dust processing}
The hydrodynamic simulations model only the gas phase of the ejecta environment. To investigate dust advection and dust destruction as well as potential dust growth in a gas, we have developed the parallelised dust-processing code \textsc{\mbox{Paperboats}} which makes use of the time- and spatially-resolved density, velocity and temperature output of the hydrodynamic simulations. In order to calculate the spatial distribution of the dust particles it makes use of the ``dusty-grid approach'' where the dust location is discretised to spatial cells in the domain and its advection is based on fluid equations. Furthermore, the dust in each cell is apportioned in different grain size bins. In a discretised time-step, the dust mass moves to other cells (due to gas streaming) or to other size bins (due to dust destruction or grain growth). The nature of the post-processing prohibits considering any feedback between the dust medium and the surrounding gas. 

We consider kinematic and thermal sputtering as well as fragmentation and vaporisation in grain-grain collisions as destruction processes. The impact of grain and plasma charges and the size-dependence (\citealt{Bocchio2014}) of the sputtering yields are taken into account. Dust growth is realised via grain sticking in grain-grain collisions as well as by re-accretion of destroyed dust material in a ``negative'' sputtering yield. The applied dust processes are described in detail in \cite{Kirchschlager2019b}. 

\section{Results}
Using our post-processing code \textsc{Paperboats} we have studied dust destruction rates in the clumpy ejecta of the SNR Cas~A as a function of initial grain sizes, dust material and gas density contrast $\chi$. We found that up to $\unit[40]{\%}$ ($\unit[30]{\%}$) of the silicate (carbon) dust mass is 	able to survive the passage of the reverse shock. The survival rates depend strongly on the initial grain size distribution, with $a\,{\sim}\,\unit[10 - 50]{nm}$ and ${\sim}\,\unit[0.5-1.5]{\textrm{\textmu} m}$ as the grain radii that show the highest surviving dust masses. The dust processing causes a rearranging of the initial grain size distribution. Our results show that grain-grain collisions and sputtering are synergistic and that grain-grain collisions have to be taken into account when dust survival rates in supernova remnants are studied.

\small  
%
\section*{Acknowledgments}   
%
 FK, FS, MJB, AB and FP were supported by European Research Council Grant SNDUST ERC-2015-AdG-694520. FP acknowledges funding by the STFC. This work was in part supported by the US~Department of Energy through the Los Alamos National Laboratory. Los Alamos National Laboratory is operated by Triad National Security, LLC, for the National
Nuclear Security Adminstration (Contract No.~89233218CNA000001).

 \begin{multicols}{1}
\bibliographystyle{aj}
 \footnotesize
\bibliography{proceedings} 
 \end{multicols}

\end{document}